\documentclass[aps,prc,reprint,groupedaddress,onecolumn]{revtex4}
\usepackage{amsmath}
\usepackage{amsfonts}
\usepackage{amssymb}
\usepackage{amscd}
\begin{document}
\title{Light-front quantum mechanics and quantum field theory}

\author{W.~N.~Polyzou}
\affiliation{Department of Physics and Astronomy, The University of
Iowa, Iowa City, IA 52242}

\thanks{Contribution to Light Cone 2019 - QCD on the light cone: from hadrons to heavy ions, 16-20 September 2019, L'\'Ecole Polytechnique, Palaiseau, France.
This research 
supported by the US Department of Energy, Office of Science, 
grant number DE-SC0016457
}
\begin{abstract}

This contribution discusses some of the advantages and
unique properties of relativistic quantum theories with kinematic
light-front symmetries.
\end{abstract}

\maketitle

\section{Introduction}

In 1939 E.P. Wigner \cite{wigner:1939} showed that a necessary and sufficient condition
for quantum observables, (probabilities, expectation values, and
ensemble averages) to have the same values in all inertial coordinate
systems is the existence of a unitary ray representation of the
component of the Poincar\'e group connected to the identity.
In order to satisfy the commutation relations
involving boost and translational generators,
\[
[K^i,P^j ] = i \delta_{ij}H, 
\]
interactions must appear in at least three of the Poincar\'e generators.

In 1949 P.A.M. Dirac \cite{dirac:1949} identified three representations of a
relativistic dynamics with the largest interaction independent
subgroups of the Poincar\'e group.  The kinematic subgroups identified
by Dirac include the Lorentz group (point-form dynamics), the
three-dimensional Euclidean group (instant-form dynamics), and the subgroup
of the Poincar\'e group that leaves a hyperplane
\[  
x^+:= x^0 + \hat{\mathbf{z}}\cdot \mathbf{x}=0 
\]
tangent to the light cone invariant (light-front dynamics).  This last
subgroup is a seven-parameter subgroup, while the Lorentz group and
three-dimensional Euclidean groups are six-parameter subgroups.
Relativistic quantum theories where there are no interactions in the
generators of this seven-parameter subgroup are called light-front
quantum theories.  Light-front representations have the smallest
number of generators that require interactions.

The seven-parameter subgroup that leaves the light-front hyperplane
invariant includes (1) a three-parameter subgroup of translations
tangent to the hyperplane (2) a three-parameter subgroup of
light-front preserving Lorentz boosts and (3) a one-parameter subgroup of
rotations about the longitudinal, $(\hat{\mathbf{z}})$, axis.  In
the $SL(2,\mathbb{C})$ representation the light-front Lorentz
transformations are
represented by the subgroup of lower triangular matrices:
\[
\Lambda_{fb}(p/m) =
\left (
\begin{array}{cc}
\sqrt{{p^+/ m}} & 0 \\
{p_\perp /\sqrt{p^+m}} & \sqrt{{m / p^+}}\\
\end{array}    
\right ) \qquad \mbox{ light-front boosts}
\]
\[
\Lambda_{fr}(\phi) =
\left (
\begin{array}{cc}
e^{i \phi/2} & 0 \\
0 & e^{-i \phi/2} \\
\end{array}    
\right )
\qquad \mbox{rotations about }\hat{\mathbf{z}}
\]
where the corresponding $4 \times 4$ Lorentz transformations are
$
\Lambda^{\mu}{}_{\nu} ={1 \over 2}\mbox{Tr}(\sigma_{\mu}\Lambda \sigma_{\nu} \Lambda^{\dagger}) .
$
The
generators of the transverse boosts and the longitudinal rotations
also satisfy the commutation relations of the two-dimensional Euclidean
Lie algebra.

The light-front Hamiltonian, $P^-$, is one of the three dynamical
generators.  It satisfies the light-front dispersion relation
\[
P^-:=
H -\hat{\mathbf{z}}\cdot \mathbf{P} =
{\mathbf{P}_{\perp}^2 + M^2 \over P^+} .
\]
One property that distinguishes a light-front dynamics
from Dirac's other forms of dynamics is that a light-front
hyperplane is not a suitable initial value surface
because it has light-like tangent vectors.

In order to compare spins of particles in different inertial frames,
it is useful to boost to a common frame where the particle's spins can
be compared.  The frame most often used is the particle's (or system's)
rest frame, but the spin defined this way depends on the boost used to
transform to the rest frame, since a boost right multiplied by a
rotation is still a boost.  The light-front spin is defined using
light-front preserving boosts:
\[
S_f^i = {1 \over 2} \sum \epsilon_{ijk} \Lambda_{fb}^{-1}({P/M})^j{}_{\mu}
\Lambda_{fb}^{-1}({P/M})^k{}_{\nu}J^{\mu \nu}
\]
where $J^{\mu\nu}$ is the angular momentum tensor, and the parameters,
$P/M$, in the boosts are operators. 
Because the light-front preserving boosts
form a subgroup, they cannot generate Wigner rotations.   This means that
light-front boosts leave the light-front magnetic quantum numbers unchanged.

One interesting property of a light-front dynamics is that the rest
frame is defined by the dynamical constraint $P^-=P^+=M$.  This means
that the conditions that determine the rest frame of free and
interacting systems do even commute.  This has dynamical implications
for the spins, because even though the light-front boost are
kinematic, the rest frame, where spins in different frames are
compared, is dynamical.

A basis for a particle of mass $m$ and spin $s$ can be constructed out of
simultaneous eigenstates of the mass, spin, and four other mutually
commuting functions of the Poincar\'e generators.  They can be taken
as the generators of translations tangent to the light-front hyperplane 
and the longitudinal component of the light-front spin:
\[
M^2, S^2, \underbrace{P^+, P^1 ,P^2}_{{\tilde{\mathbf{P}}}}, S^3_{f} . 
\]
The single-particle basis vectors are 
\[
\vert (m,s) \tilde{p},\mu \rangle.
\]
Lorentz covariance of the four-momentum and angular-momentum tensors determine
both the transformation properties and spectra of these observables.
The resulting transformation property of a mass $m$ spin $s$
irreducible representation of the Poincar\'e group is:
\[
U(\Lambda ,a) \vert (m,s) \tilde{p},\mu \rangle =
e^{i p' \cdot a}\sum_\nu
\vert (m,s) \tilde{p}',\nu \rangle
\sqrt{{p^{+\prime} \over p^+}}
D^s_{\nu \mu}[\Lambda_{fb}^{-1}(p'/m) \Lambda \Lambda_{fb}(p/m)]
\]  
where $\Lambda_{fb}(p/m)$ are light-front boosts from the rest frame to
$\tilde{\mathbf{p}}$ and $p'=\Lambda p$.  

While the light-front spin is invariant with respect to light-front
boosts, it undergoes Wigner rotations under ordinary rotations.  More
importantly, the light-front Wigner rotation of a rotation is not the
rotation.  What this means is that if a system of non-interacting
particles is rotated, the spin of each particle will rotate with a
different angle that depends on the particle's momentum.  This means
that light front-spins cannot be added with ordinary $SU(2)$
Clebsch-Gordan coefficients.

In order to add light-front spins for systems of free particles it is
necessary to boost to the system rest frame, transform the light-front
spins to canonical spins, add the canonical spins and orbital angular
momenta, then boost the back to the original frame with a light-front
boost.  The coefficients of the resulting unitary transformation
are the Clebsch-Gordan coefficients for the Poincar\'e group in a light-front
basis.  The rotations that transform the light-front spins to canonical
spins, are called Melosh rotations. They involve a light-front boost
followed by an inverse rotationless (canonical) boost,
\[
R_M(p/m) = \Lambda^{-1}_{cb}(p/m) \Lambda_{fb}(p/m) .
\]

It is more convenient to treat systems of particles in a many-body
basis where the internal degrees of freedom are invariant with respect
to light-front boosts.  These variables include the total light-front
momenta, light-front momentum fractions, the transverse single-particle
three momenta transformed to the kinematic rest frame with light-front
boosts, and the light-front magnetic quantum numbers.  These variables
have the property that only the total light-front momentum transforms
under light-front boosts; all of the other variables are invariant.

Dynamical light-front models can be constructed by adding
kinematically invariant interactions to the non-interacting
light-front Hamiltonian
\[
P^- \to (\mathbf{P}^2_{\perp} + M_0^2 + V )/P^+
\]
of a many-particle system.  Allowable interactions must preserve the
spectral condition and result in a self-adjoint $P^-$.  Rotational
covariance is an additional non-trivial dynamical constraint.  These
conditions can be realized in few-body models.  When $P^-$ is
self-adjoint the dynamics is well-defined and given by a one-parameter
unitary group, so in this case the self-adjointness of the light-front
Hamiltonian ensures that there is no ambiguity associated with the bad
initial value surface.

The Light-front representation has advantages for computing current
matrix elements.  There are several reasons for this.  First, since
the boosts are kinematic, boosts of the initial and final states can
be computed by applying the inverse boost to arguments of the
light-front wave function, which is normally expressed in a
non-interacting multi-particle basis.  In addition, because the
light-front boosts form a subgroup, the light-front spins in the
initial and final states of the the current matrix elements are frame
independent.  If the momentum transfer is space-like, the orientation
of the light front can be adjusted so all current matrix elements can
be expressed in terms of matrix elements of $I^+(0)$, and these matrix
elements are invariant under light front-boosts.  Another unique
feature of the light-front representation is that for one-body
(impulse) current operators the momentum transferred to the system is
the same as the momentum transferred to the constituents in all frames
related by light-front boosts.  This is not the case in Dirac's
instant or point-form dynamics, however in the light-front case for
spin greater than $1/2$ there are more current matrix elements than
there are independent form factors.  The additional current matrix elements are
related by dynamical rotational covariance.

A true covariant current in a dynamical model cannot be a one-body
operator and satisfy current covariance and current conservation.
Impulse approximations can be made by assuming that the two-body
current vanishes on a set of independent matrix elements.  The
remaining matrix elements can be computed by imposing rotational
covariance.  This is not as satisfactory as having a covariant current
operator that can be used with different initial and final states.

One feature of Dirac's different representations of the dynamics is
that they are all scattering equivalent (related by an $S$-matrix
preserving unitary transformation).  This means that given generators
in one representation it is possible to find equivalent generators in
any other representation.  This equivalence suggests that it is useful to
exploit the advantages of the light-front representation.

The interest in the light-front representation is motivated by
properties of the light-front formulation of quantum field theory.
For free fields, light-front fields are constructed by changing
variables from three momenta to the three light-front components of
the four momentum.  One feature of the light-front field is that a
canonical momentum field is not needed to separate the creation and
annihilation operators: 
\[
a(\tilde{\mathbf{p}}) = \sqrt{p^+\over 2}\theta (p^+)
\phi (x^+=0,p^+,\mathbf{p}_{\perp})
\qquad
a^{\dagger} (\tilde{\mathbf{p}}) = \sqrt{p^+\over 2}\theta (p^+)
\phi (x^+=0,-p^+,\mathbf{p}_{\perp}) .
\]
What this means is that any operator that
commutes with the field restricted to a light front must be a constant
multiple of the identity.  This means that free fields restricted to
the light front are irreducible.
%

A related property of the free fields restricted to the light front
is that the fields are independent of the mass.  For canonical
fields, the canonical transformation that changes masses cannot be
implemented by a unitary transformation \cite{haag:1955},
while fields restricted to a
light front are trivially related by a unitary transformation
\cite{leutwyler:1970}.

Another property of light-front field theory follows because $P^+$ is
a kinematic operator satisfying the spectral condition $P^+ \geq 0$.
Interactions that preserve the kinematic symmetries must commute with
$P^+$.  This means that $V \vert 0 \rangle $ is an eigenstate of $P^+$
with eigenvalue 0.  It follows that
\[
\langle 0 \vert V V \vert 0 \rangle =
\langle 0 \vert V \vert 0 \rangle \langle 0 \vert V \vert 0 \rangle,  
\]
because there can be no contribution from states with absolutely
continuous spectrum of $P^+$ since $P_i^+=0$ is a set of measure 0.
This requires 
$
V\vert 0 \rangle = \vert 0 \rangle \langle 0 \vert V \vert 0 \rangle .
$
Invariance of the vacuum can be realized by a simple constant
renormalization of $P^-$, with the vacuum remaining the Fock vacuum.

The role of the Fock-vacuum in light-front dynamics can be best
understood by considering Noether's theorem on the light front.  The
Poincar\'e invariance of the action leads conserved Noether currents.
Integrating the ``$+$'' components of these currents over the light
front, assuming no contributions from the boundaries, means that all
10 Poincar\'e generators are independent of $x^+$.  The expression for
these generators involves fields on the light front and derivatives of
these fields.  While the fields and derivatives tangent to the light
front are all in the irreducible light-front algebra, the derivatives
of the fields normal to the light front are unconstrained.

This ambiguity is related to the problem with the inadequacy of the
light front as an initial value surface.  This does not mean that the
irreducibility cannot be exploited.  It means the additional
information is needed to define the derivatives off of the light-front
in terms of the irreducible algebra on the light front.  For a scalar
field theory all of the normal derivatives cancel.  In this case all
the generators can be expressed in terms of the operators in the
irreducible light front algebra.  The dynamical generators have the form
\[ 
K^3=
\int_{x^+=0} d\tilde{\mathbf{x}} : \left (4{\partial \phi(x) \over \partial x^-}
    {\partial \phi(x) \over \partial x^-} x^- -
   ( \pmb{\nabla}_{\perp} \phi(x) \cdot \pmb{\nabla}_{\perp} \phi(x)
+m^2\phi(x)\phi(x) +V (\phi)) x^+ \right ):
\]
\[
P^- =  \int_{x^+=0} d\tilde{\mathbf{x}} :
\left ( \pmb{\nabla}_{\perp} \phi(x) \cdot \pmb{\nabla}_{\perp} \phi(x)
+m^2\phi(x)\phi(x) +V (\phi) \right ):
\]
\[
F^i =
 \int_{x^+=0} d\tilde{\mathbf{x}} :
 \left (
(\pmb{\nabla}_{\perp} \phi(x) \cdot \pmb{\nabla}_{\perp} \phi(x)
+m^2\phi(x)\phi(x) +V (\phi))x^i + 2 
x^- {\partial \phi(x) \over \partial x^-} {\partial \phi(x) \over \partial x^i}
)\right ):.
\]
$K^3$ has a dynamical component, but it is multiplied by $x^+$ which
vanishes on the light front.
Iterating the light-front Heisenberg field
equations, assuming that the commutation relations of the fields
on the light front agree with the free-field commutation relations,
\[
[\phi (\tilde{\mathbf{x}},x^+=0), \phi (\tilde{\mathbf{y}},y^+=0)]
= i {\pi \over 4} \delta (\mathbf{x}_{\perp} -\mathbf{y}_{\perp})\epsilon (x^-- y^-) 
\]
results in a series in $x^+$ with coefficients that involve fields and
commutators of fields restricted to the light front.

The expressions for the dynamical generators have the usual diseases
in that products of fields at the same point are ill-defined.  For the
case of free fields the light-front Heisenberg equations are linear
and can be solved, resulting in the correct expression for the field
operator of mass $m$ in the light-front representation. Vacuum
expectation values of products of these fields, no longer restricted to the
light front, in the trivial vacuum
give the Wightman functions of the free field theory.  These are
moments of the dynamical (mass-dependent) vacuum.  While on one hand
this result is trivial, free fields with different masses live on
different Hilbert spaces with different vacuua. This shows that the
correct result can still be obtained by using the trivial vacuum and the
light-front field algebra with different interactions.

In an interacting field theory additional information is necessary to
define the theory.  This involves finding a non-perturbative way to
define products of fields at the same point.  Normally non-trivial
vacuua are due to the all creation operator terms in the Hamiltonian.
In the case of a $\phi^4(x)$ theory the creation operator part of the
light-front $P^-$ has a term with general form
\[
\int {\theta (p^+) \delta (p^+) dp^+\over (p^+)^2 \prod \xi_i}
\prod d\mathbf{p}_{i\perp} d\xi_i \delta (\sum \mathbf{p}_{i\perp})
\delta (\sum \xi_i -1) \times
\]
\[
a^{\dagger}(\xi_1 p^+, \mathbf{p}_{\perp 1})
a^{\dagger}(\xi_2 p^+, \mathbf{p}_{\perp 2})
a^{\dagger}(\xi_3 p^+, \mathbf{p}_{\perp 3})
a^{\dagger}(\xi_4 p^+, \mathbf{p}_{\perp 4}). 
\]
While formally the $\delta (p^+)$ suggests that the vacuum should
remain trivial, this contribution is ill-defined (and very singular)
at $p^+=0$.  The fate of the trivial vacuum depends on what replaces
this operator.  The singularities at $p^+=0$ are not independent of
the ultraviolet singularities of the theory, since they are
transformed into each other as the orientation of the light front is
rotated.  Both need to be addressed in order to construct a
self-adjoint $P^-$.

When the theory has infrared singularities, additional contributions
concentrated at $P^+=0$ may be required to maintain equivalence with
the canonical theory.  These affect the interpretation of the
light-front vacuum.  These zero-modes are required in perturbation theory.
While these zero modes disappear in a theory with cutoffs; it is not clear
if there is a way to recover them as the cutoff is removed.
or if they have to be put in by hand.

\end{document}